\documentclass[compsoc,conference,a4paper,10pt,times]{IEEEtran}

\usepackage{caption, subcaption, graphicx}
\usepackage{multirow, xspace, booktabs}
\usepackage{svg}
\usepackage{float}
\usepackage{pifont}
\usepackage{algorithm}
\usepackage{algpseudocode}
\usepackage{comment}
\usepackage{fancyhdr}
\usepackage{array}
\usepackage{tikz}
\pdfoutput=1

\usepackage{cite}
\usepackage{amsmath,amssymb,amsfonts}
\usepackage{textcomp}
\usepackage{bmpsize}
\usepackage{xcolor}
\usepackage{lipsum}
\usepackage[colorlinks=true,urlcolor=black]{hyperref}
\def\BibTeX{{\rm B\kern-.05em{\sc i\kern-.025em b}\kern-.08em
    T\kern-.1667em\lower.7ex\hbox{E}\kern-.125emX}}

\usepackage[normalem]{ulem}
\useunder{\uline}{\ul}{}

\usepackage{cleveref}
\crefformat{section}{\S#2#1#3}
\crefformat{subsection}{\S#2#1#3}
\crefformat{subsubsection}{\S#2#1#3}

\newif{\ifanonymous}
\anonymoustrue{} 

\newcommand{\eg}{e.g.,\ }
\newcommand{\etal}{et al.\xspace}
\newcommand{\ie}{i.e.,\ }
\newcommand{\cf}{{\em cf.}\ }

\newcommand{\parait}[1]{\vspace{.05in}{\it \noindent #1 }}
\newcommand{\para}[1]{\vspace{.05in}{\bf \noindent #1 }}
\newcommand{\sellers}{\texttt{sellers.json}\xspace}
\newcommand{\ads}{\texttt{ads.txt}\xspace}

\newcommand{\brand}{{advertiser}\xspace}
\newcommand{\brands}{{advertisers}\xspace}

\newcommand{\adx}{{ad-network}\xspace}
\newcommand{\adxs}{{ad-networks}\xspace}

\newcommand{\vpub}{{victim publisher}\xspace}
\newcommand{\vpubs}{{victim publishers}\xspace}

\newcommand{\academics}{{academic researchers}\xspace}

\newcommand{\cma}{{activists}\xspace}

\newcommand{\probpools}{{\em probdomains}$_{vp}$\xspace}

\newcommand{\darkpools}{{\em pools}$_{ad}$\xspace}
\newcommand{\partnerdarkpools}{{\em partnerpools}$_{ad}$\xspace}

\newcommand{\quoteResponse}[1]{``\textit{#1}''\xspace}

\begin{document}

\title{Towards Multi-Stakeholder Vulnerability Notifications \\in the Ad-Tech Supply Chain}

\author{
{\rm Yash Vekaria}\\
University of California, Davis\\
yvekaria@ucdavis.edu
\and
{\rm Rishab Nithyanand}\\
University of Iowa\\
rishab-nithyanand@uiowa.edu
\and
{\rm Zubair Shafiq}\\
University of California, Davis\\
zubair@ucdavis.edu
}

\maketitle


\begin{abstract}
Online advertising relies on a complex and opaque supply chain that involves multiple stakeholders, including advertisers, publishers, and ad-networks, each with distinct and sometimes conflicting incentives.
Recent research has demonstrated the existence of ad-tech supply chain vulnerabilities such as dark pooling, where low-quality publishers bundle their ad inventory with higher-quality ones to mislead advertisers.
We investigate the effectiveness of vulnerability notification campaigns aimed at mitigating dark pooling.
Prior research on vulnerability notifications have primarily explored single-stakeholder contexts, leaving multi-stakeholder scenarios understudied. 
There is limited attention to complex multi-stakeholder supply chain ecosystems such as ad-tech supply chain, where resolving vulnerabilities often requires coordinated action across entities with misaligned incentives and interdependent roles. 
We address this gap by implementing the first online advertising supply chain vulnerability notification pipeline to systematically evaluate the responsiveness of various stakeholders in ad-tech supply chain, including publishers, ad-networks, and advertisers to vulnerability notifications by academics and activists. 
Our nine-month long automated multi-stakeholder notification study shows that notifications are an effective method for reducing dark pooling vulnerabilities in the online advertising ecosystem, especially when targeted towards ad-networks. 
Further, the sender reputation does not impact responses to notifications from activists and academics in a statistically different way.
Overall, our research fosters industry-scale solution to combat ad inventory fraud and fosters future research on feasibility of multi-stakeholder vulnerability notifications in other supply chain ecosystems.
\end{abstract}


\section{Introduction}

\para{Fraud is rampant in the online advertising supply chain.}
In 2023, marketers globally spent more than 300 billion dollars on online advertising \cite{ad-spend-bechmarks-2023}. 
But nearly a quarter of this ad spend is lost to ad fraud \cite{juniper}.
Ad fraud generally takes one of three forms: inorganic or fraudulent interactions with ad content \cite{pearce-ccs14, kshetri-sp10, springborn-security13}, manipulating conversion attribution \cite{chachra-imc2015, stone-imc2011}, and ad inventory laundering \cite{vekaria2023inventory, papadogiannakis2023funds, kline2022placement, inventorymisrepresentation}. 
While the vulnerabilities exploited by each form of ad fraud are multifaceted, they generally exploit the complexity and opacity of the online advertising supply chain \cite{supplychaintransparency}. 
{\em This paper focuses on an emerging type of ad inventory laundering called dark pooling}. 
At a high-level, dark pooling allows low-quality (\ie brand-unsafe) publishers make their ad inventory indistinguishable, to advertisers or brands, from the ad inventory of higher-quality publishers. 
Recent work \cite{vekaria2023inventory} has shown that dark pooling is widespread and helps fund low-quality websites known for publishing misinformation and other brand-unsafe content.

\para{Current dark pooling mitigation strategies are ineffective.}
Solutions to prevent the laundering of low-quality ad inventory facilitated by dark pooling generally take the form of new standards for improving transparency and facilitating inventory verification in the online advertising supply chain. 
For instance, the Interactive Advertising Bureau (IAB) introduced standards such as {\tt ads.txt} \cite{adstxt-specification}, {\tt sellers.json} \cite{sellersjson-specification}, and the Supply Chain Object \cite{openrtb-guidelines} to help mitigate this type of ad fraud. 
Unfortunately, prior research consistently shows that entities do not comply with these standards \cite{bashir2019longitudinal, papadogiannakis2025welcome, vekaria2023inventory, tingleff2019three, pastor2020establishing}. 
Consequently, these measures fail to mitigate, or even detect, ad inventory fraud. 
Further, given the largely self-regulated nature of the ad-tech ecosystem, we cannot expect significant improvements in compliance rates \cite{zukina2012accountability, culnan2000protecting, hirsch2010law}. 
Therefore, it is important to explore alternative approaches to mitigate ad inventory fraud.
{\em This paper examines the effectiveness of a different tactic, previously unused in the ad-tech supply chain: a vulnerability notification campaign}. 
Specifically, we examine the effectiveness of notification campaigns in mitigating dark pooling.

\para{Findings from prior vulnerability notification research are unlikely to carry over to the online advertising supply chain.}
In context of the Web, researchers have conducted many studies on how different notification campaign strategies affect the resolution of various vulnerabilities observed in online services \cite{stock2016hey,li2016you,li2016remedying,cetin2016understanding,cetin2017make,ccetin2019tell,maass2021snail,maass2021effective}. 
These prior works primarily focus on single-stakeholder scenarios, where vulnerability notifications are sent to a single entity that is capable of resolving it (\eg a website operator).
A few studies have examined multi-stakeholder, but typically in contexts where the vulnerability (1) is associated with a single stakeholder (\eg WHOIS and IP WHOIS notified about non-secure DNS vulnerability associated solely with nameserver operator~\cite{cetin2017make}) or (2) does not require coordinated effort from two or more notified stakeholders to fix the vulnerability (\eg either hosting providers or CERTs can independently perform remediations to handle the notification about vulnerable websites, without a need to coordinate~\cite{stock2016hey})
However, there remains limited understanding of the feasibility and effectiveness of vulnerability notifications in complex and opaque multi-stakeholder ecosystems such as ad-tech supply chain, where
(1) vulnerability could exist at multiple stakeholders.
(2) vulnerability resolution may require cooperation from multiple independent entities (\eg publishers may need to coordinate with ad-networks to fix the dark pooling vulnerability);
(3) incentives of the independent entities may be different and at odds with each other.
Designing notification strategies for such ecosystems requires incorporating these interdependencies and incentive misalignments into the study design.
{\em Our work fills this gap by studying multi-stakeholder ad-tech supply chain ecosystem for the first time.}

\para{Developing a vulnerability notification campaign for the ad-tech supply chain is non-trivial.}
Our research addresses several challenges related to the opacity of the multi-stakeholder ad-tech supply chain.
First, we need to automate the detection of dark pooling vulnerabilities to conduct notifications at a reasonable scale. 
We address this challenge using methods from prior research \cite{vekaria2023inventory} to identify dark pooling and the specific entities involved, such as publishers, ad-networks, and advertisers.
Second, we must tailor our notifications to make them understandable and useful to each type of stakeholder. 
We accomplish this by designing notifications that provide entity-specific information, including high-level descriptions of the detected dark pooling vulnerability, technical evidence, and potential remedial actions.
Finally, we need to assess the effectiveness of our notifications with statistical rigor and correctly attribute vulnerability resolutions to different stakeholders. 
We achieve this by conducting our study over a nine-month period and in multiple phases, each focused on inferring the actions of one type of stakeholder.

\para{Our study design and findings advance both notification research and ad fraud research.} 
We first design a multi-phase email-based notification campaign for mitigating dark pooling in the multi-stakeholder ad-tech supply chain (\Cref{sec:methodology}). 
This design provides a template for future fragmented multi-stakeholder notification studies. 
Next, we operationalize this design to identify and notify 2.9K entities (1.7K publishers, 644 ad-networks, and 635 advertisers) involved in and affected by dark pooling vulnerabilities. 
We analyze their responses to answer the following research questions:

\noindent $\bullet$ {\em RQ1. What are the differences in stakeholder attitudes and responses towards dark pooling vulnerability notifications? (\Cref{sec:engagement})} 
We conduct a thematic analysis of the stakeholder responses to our notifications. Our approach uncovers three response themes: (1) effort towards resolution; (2) demonstration of concern or awareness; and (3) lack of trust in our data or lack of resources for resolution.

\noindent $\bullet$ {\em RQ2. How does the choice of notification recipient influence the resolution of dark pooling vulnerabilities? (\Cref{sec:impact:recipient})}
We measure the rates of dark pooling vulnerability resolution when notifications are sent to publishers, ad-networks, and advertisers. We find that sending notifications to ad-networks has the highest influence on dark pooling vulnerability resolution with 81.6\% of notified entities remediating dark pools in their network and 76.9\% in partner ad-networks. Publishers demonstrate the least effectiveness to notifications, however, with the highest effect size.

\noindent $\bullet$ {\em RQ3. How does the notification sender influence the resolution of dark pooling vulnerabilities? (\Cref{sec:impact:sender})}
We compare the efficacy of notifications sent from academics with those from activists with a long history of public advocacy in addressing online advertising vulnerabilities \cite{ferraz2021sleeping, braun2019activism, checkmyads}. We find that notifications sent as activists are generally as effective as those from academics. The only exception is when the notifications are sent to publishers. In this case, activists have a statistically significantly stronger positive effect on the resolution of dark pools.

\parait{All together, our design, analysis methods, and results provide crucial insights into the applicability of vulnerability notifications in complex, multi-stakeholder environments.} 

\section{Background} \label{sec:background}

\begin{figure}
    \centering
    \includegraphics[width=\linewidth]{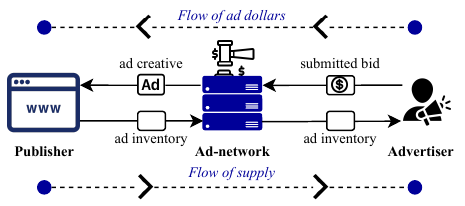} 
    \caption{Stakeholders in the ad-tech supply chain}
    \label{fig:teaser}
\end{figure}

\subsection{The online advertising supply chain} \label{sec:background:supplychain}

\para{Stakeholders in online advertising.}
The online advertising supply chain involves three main types of stakeholders:
(1) publishers (\ie websites), who are the source of the ad inventory; 
(2) ad-networks (\ie supply-side platforms and ad-exchanges) that facilitate a real-time bidding marketplace for ad inventory; and 
(3) advertisers (\ie brands), who buy ad inventory from publishers to display their ads to users.

\para{How the online advertising supply chain works.}
Publishers, ad-networks, and advertisers collaborate to create the ad supply chain. 
When a user visits a publisher's website, the ad inventory associated with that visit is auctioned off at an ad-network (possibly by other ad-network intermediaries called supply-side platforms).
Advertisers then make bids on the auctioned ad inventory available at the ad-network (with the help of brokers called demand-side platforms). 
Finally, the ad-network places the ad from the winning advertiser (e.g., the highest bid) on the publisher’s website. The advertiser transfers the payment to the ad-network (possibly through intermediary networks), and deposits a portion of the billed impressions into the publisher’s account (possibly through intermediaries).
A simplified version of this process is illustrated in \Cref{fig:teaser}.

\para{Control and incentives in the online advertising supply chain.}
Each ad-tech stakeholder has different levels of control and different incentives.

\parait{Publishers.}
They aim to get the highest bids for their ad inventory by participating in auctions at large ad-networks, enrolling with multiple ad-networks, and showcasing the value of their user base to advertisers. They do not control other parts of the supply chain. 

\parait{Ad-networks.}
They earn a fraction of each winning bid at ad inventory auctions. They aim to maximize the number of clients (\ie publishers and other ad-networks) participating in their auctions and ad inventory sold through their platforms. 
As facilitators of the ad inventory auction, ad-networks have significant control and visibility into both ends of the supply chain. However, this is diminished when multiple non-collaborative ad-networks participate in a transaction (this is the common case).

\parait{Advertisers.}
Their goal is to ensure their ads reach the right audience. Despite funding the entire supply chain with their advertising dollars, they cannot verify if their ads are shown to the right users and on brand-safe publishers or not.

\para{Challenges with the multi-stakeholder supply chain.}
One consequence of this decentralized multi-stakeholder scenario is that vulnerabilities can exist at multiple stakeholders and resolution is rarely possible without collaboration between multiple entities, each with their own controls and operating incentives. 
This interdependence and necessary trust, despite conflicting objectives, between different stakeholders makes the online advertising supply chain a uniquely interesting case-study for assessing the effectiveness of vulnerability notifications.

\subsection{Dark pooling vulnerabilities} \label{sec:background:pooling}

\para{Seller ID pooling in online advertising.}
A {\em seller ID} is assigned to an ad inventory seller’s account when they establish a relationship with an ad-network. 
This seller can be a publisher or another ad-network. 
Pooling is a strategy used to simplify inventory management for organizationally-related publishers. 
Sellers owning multiple publishers (\eg parent organizations with several websites) manage their inventory through a single seller account on an ad-network and receive just one seller ID for the account. 
This practice allows for efficient operation and inventory management, but current transparency standards make it possible to mask the exact inventory source (publisher) when it belongs to a set of pooled websites sharing the same seller ID. 
This trade-off is generally accepted due to the assumed similar reputations of publishers owned by the same entity.

\begin{figure}
    \centering
    \includegraphics[width=\linewidth]{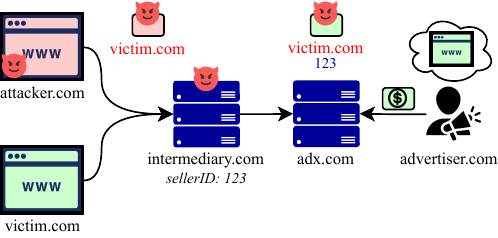} 
    \caption{Our threat model representing ad-tech supply chain vulnerability of dark pooling.}
    \label{fig:threat-model}
    \vspace{-2mm}
\end{figure}

\para{How pools become dark?} 
A pool becomes a {\em dark pool} when the publishers sharing a single seller ID are {\em organizationally-unrelated and have different reputations} \cite{vekaria2023inventory}. 
Dark pooling allows low-quality publishers to disguise their inventory as high-quality, leading advertisers to unintentionally purchase low-quality, brand-unsafe inventory. 
Malicious publishers or ad-networks can facilitate dark pooling. 
Simple examples of both types of dark pooling vulnerabilities are shown in \Cref{fig:threat-model} and explained below. 
More comprehensive explanations can be found in Vekaria \etal \cite{vekaria2023inventory} and Papadogiannakis \etal \cite{papadogiannakis2025welcome}.
The remainder of our paper is focused on using notification campaigns to capture the motivations, control, and capabilities of each stakeholder in addressing inventory fraud due to dark pooling.

\parait{Dark pooling facilitated by publishers.}
Recall that publishers aim to join large ad-networks to maximize revenue, which is challenging for low-quality publishers.
Dark pooling allows these publishers to misrepresent their existing relationships in order to gain access to large ad-networks. 
For example, a low-quality publisher ({\tt attacker.com}) manipulates its \ads file to \textit{intentionally} hide or misrepresent its ad inventory by including entries from \ads of some reputable publisher ({\tt victim.com}), making attacker's inventory appear as though it belongs to a reputable publisher. Next, they change the inventory source to {\tt victim.com} in future bid requests. 
An unwitting brand ({\tt advertiser.com}) might then bid on this inventory, incorrectly believing that it belongs to {\tt victim.com} (even after using {\tt victim.com}'s \ads file for source verification).
Although {\tt attacker.com} may not immediately financially benefit from the inventory sold by this manipulation (payments go to {\tt victim.com}'s account), it is useful for demonstrating a higher value for the ad inventory on its website to other ad-networks.
Such manipulation is possible because of: (1) misrepresentations, incompleteness, and inaccuracies within \ads and \sellers files made available by publishers and ad-networks, respectively; and (2) poor verification procedures by ad-networks which are incentivized to maximize the sale of ad inventory. Vekaria \etal \cite{vekaria2023inventory} discuss these manipuations in detail.

\parait{Dark pooling facilitated by ad-networks.}
Intermediary ad-networks ({\tt intermediary.com}) provide `inventory management' services to smaller publishers unable to manage their own inventory with in-house teams. 
Dark pooling occurs when these intermediaries share a single seller ID among clients (\ie publishers) with different reputations. 
In this case, ad inventory owned by {\tt victim.com} and {\tt attacker.com} auctioned through the {\tt intermediary.com}'s network appears to an advertiser as owned by {\tt intermediary.com} and can be made indistinguishable (for e.g., {\tt attacker.com} misrepresents its inventory source as {\tt victim.com}).
Ad-networks are not incentivized to prevent dark pooling as it would reduce the volume of inventory traded.

\noindent In summary, our dark pooling threat model may constitute following scenarios:\\
\ding{182} Ad-network bundles ad inventory from different publishers and sells them to buyers using a common sellerID owned by the ad-network. Given the challenges with source verification, ad-network (knowingly or unknowingly) sells misrepresented inventory (i.e., a problematic publisher (mis)represented as victim[.]com).\\
\ding{183} Problematic and victim publishers may (not) collude:\\ 
$\bullet$ If they do not collude: Problematic publishers may still improve their reputation by using ad-network assigned victim publishers’ IDs (e.g., to show that they work with reputable ad-networks to convince more advertisers to bid or gain access to new ad-networks) as a result of pooling. \\
$\bullet$ If they collude: Problematic and victim publishers could have some sort of a revenue sharing model where problematic publishers share revenue generated by using victim's seller ID with the victim publisher. However, if an advertiser detects low ROI or ad-misplacements associated with certain pooled seller ID, they may block it -- hurting victim publisher's long term revenue.

\para{Industry efforts to mitigate dark pooling.}
The ad-tech ecosystem has introduced several standards to increase supply chain transparency and mitigate inventory fraud.
Notable among these are the \ads standard \cite{adstxt-specification}, \sellers standard \cite{sellersjson-specification}, and the RTB (real-time bidding) SCO (Supply Chain Object) \cite{openrtb-guidelines} all introduced by the Interactive Advertising Bureau. 

\parait{The \ads standard.}
Each publisher participating in RTB is expected to implement and maintain an \ads file~\cite{adstxt-specification} at the root of their domain. 
This file lists all \adx seller domains (and seller IDs assigned to the publisher by the corresponding \adx) that are authorized to sell or resell the publisher's inventory. 

\parait{The \sellers standard.}
Ad-networks must maintain a JSON file -- \sellers~\cite{sellersjson-specification} at their domain’s root. 
This file details all publishers and intermediary ad-networks they work with, along with corresponding domains and \adx-assigned seller IDs.

\parait{The Supply Chain Object.}
SCO is a field in the RTB object that each seller in the supply chain populates using information associated with their inventory source to provide transparency about involved entities. 
Buyers validate the supply chain in real-time using \ads, \sellers, and SCO, then decide whether to bid on an ad slot. 

\noindent Unfortunately, prior work has shown that these standards have relatively low adoption rates and serious misrepresentation issues~\cite{vekaria2023inventory,papadogiannakis2023funds,papadogiannakis2025welcome,bashir2019longitudinal,olejnik2018enhancing}. 
Moreover, lack of enforcement makes them unsuitable for preventing inventory fraud by itself.

\para{Related academic research.}
Recent academic research has focused on measuring the prevalence of inventory fraud. Notably, Kline \etal \cite{kline2022placement} illustrated several supply chain ad attribution exploits which allowed malicious publishers to inflate their ad revenue by making their inventory appear premium.
Papadogiannakis \etal \cite{papadogiannakis2023funds, papadogiannakis2025welcome, papadogiannakis2024ad} showed how problematic websites exploited supply chain complexities to monetize their website using hidden content, pooling, and identity masquerading.
Most closely related to this work, Vekaria \etal \cite{vekaria2023inventory} provided methods to identify dark pooling, measured its prevalence, and showed how it monetizes misinformation publishers.

\subsection{Vulnerability notification campaigns} \label{sec:background:notifications}

\para{Notifications of security and privacy vulnerabilities.}
Researchers have studied the effectiveness of notifications for warning server operators about unintended abuse of their infrastructure for malicious purposes \cite{cetin2016understanding, vasek2012malware}. 
They have also examined how well these notifications work for fixing various security vulnerabilities, such as HTTPS misconfigurations \cite{zeng2019fixing}, issues in version control system (VCS) repositories \cite{maass2021snail, stock2018didn}, firewall omissions for IPv6 services \cite{li2016you}, DDoS amplification vulnerabilities \cite{kuhrer2014exit, li2016you}, Heartbleed \cite{durumeric2014matter}, cross-site scripting (XSS) bugs \cite{stock2016hey}, and others \cite{maass2021best, maass2021effective, ccetin2018let, ccetin2019tell, cetin2017make,li2016remedying, canali2013role}. 
More recent, related research has studied the remedial effects of privacy non-compliance notifications~\cite{stover2023website, soussi2020feasibility, utz2023comparing, maass2021best}. 
Despite extensive research on notification campaigns, most studies either focus on single-stakeholder scenarios or multi-stakeholder ones where different stakeholders are notified about a vulnerability that exist at a single stakeholder~\cite{cetin2017make} or can be fixed by a single stakeholder~\cite{stock2016hey} or that has clear aligned incentives for all stakeholders~\cite{li2016you}. Ours is the first to evaluate notification campaigns in the fragmented multi-stakeholder ad-tech supply chain where incentives, responsibilities, vulnerabilities and fixes need cooperation and effort from all stakeholders.

\para{Notification campaigns in the online advertising supply chain.}
Although previous research has extensively studied ad-tech supply chain vulnerabilities, it has not examined the effectiveness of large-scale notifications in remedying these vulnerabilities.
The work of Vekaria et al. \cite{vekaria2023inventory} is closest to this study; they conducted a small-scale disclosure of dark pooling vulnerabilities to 55 reputable brands, treating it as a single-stakeholder issue and ignoring the fragmented multi-stakeholder aspect of the ad-tech supply chain.
In contrast, we build an effective and large-scale multi-stakeholder vulnerability notification framework for the online advertising supply chain. 
By focusing on dark pooling, we study the influence and attitudes of each stakeholder (publishers, ad-networks, and advertisers) towards notifications sent from academic researchers and a well-known ad-tech watchdog group, CheckMyAds \cite{checkmyads} (i.e., activists). 
Our automated framework for detecting vulnerabilities and sending notifications is publicly available at {$<$URL blinded for review$>$}, as is the (non-email) data that is the basis of this paper.

\para{Challenges with notification delivery mechanisms.}
Notification studies often involve the large-scale detection and notification of vulnerabilities \cite{durumeric2013zmap, cui2011reflections, bajpai2019dagstuhl}. 
A common challenge is reliably delivering notifications to the intended recipients. Often, notifications go undelivered due to the lack of publicly known communication channels.
Previous research has explored various non-automated notification channels, such as telephone, contact forms on websites, social media accounts, physical mail, and manually identified contact email addresses \cite{stock2018didn, maass2021effective, maass2021snail}. 
These manual approaches achieve high delivery success but require significant manual effort, making them infeasible for large-scale studies like ours, which focuses on the globally distributed multi-stakeholder ad-tech supply chain.
To address this challenge, researchers have used automated alternatives for notifications, such as targeting generic email addresses, or obtaining addresses from hosting providers \cite{cetin2016understanding}, WHOIS contact information \cite{zeng2019fixing, cetin2016understanding, li2016you, cetin2017make, stock2018didn, durumeric2014matter, stock2016hey, pouryousef2020extortion}, trusted third parties like CERTs or clearing houses \cite{stock2016hey, li2016you, cetin2017make, kuhrer2014exit}, and DNS operators \cite{cetin2017make}. 
However, these email-based automated approaches often suffer from low delivery success rates due to outdated contact details \cite{stock2016hey, cetin2017make, cetin2016understanding}, spam filters \cite{stock2016hey, stock2018didn}, or incorrect recipients who do not forward the email to the responsible entity \cite{li2016you}. Additional challenges such as lack of trust in senders also emerge \cite{ccetin2018let, ccetin2019tell, stock2018didn, zeng2019fixing}.
In general, prior works show a trade-off between scale and notification deliverability.

\section{Research Methods} \label{sec:methodology}

\subsection{Identifying dark pooling vulnerabilities} \label{sec:methodology:pooling}

\para{Curating problematic publishers.}
Problematic publishers are more than twice as likely to be pooled compared to other publishers \cite{vekaria2023inventory}. Therefore, we focus on this subset of publishers. Table~\ref{tab:problematic-dataset} lists the seven categories of problematic publishers in our dataset.
We started by collecting a list of publishers already classified as problematic by prior research. Since identifying these publishers is not our main goal, we used these existing classifications. First, we removed duplicates and checked if the publishers' websites were functional. We discarded 79,938 publishers that were either non-functional or had parked domains.
Next, we used an advertising filter list \cite{easylist} to see if the remaining 12,875 publishers had advertising. As expected, many problematic publishers (like typosquatted publishers) did not have ads. We found at least one advertising request on 1,125 of these publishers.
To avoid including legitimate publishers and confirm the presence of ads, we manually checked the publisher content. We kept 684 problematic publishers that had one or more display ads on their homepage. We also included one random subpage from these 684 publishers and from the remaining 441 publishers if the subpage served display ads.
Our final dataset contains 1,478 URLs across 1,007 publishers.

\begin{table}[t]
\centering
\footnotesize
{
\begin{tabular}{p{.72in}p{1.6in}p{.5in}}
\toprule
{\bf Category} & {\bf Dataset source (\# publishers)} & {\bf \#\hspace{1mm}retained publishers} \\
\midrule
Misinformation/ & \cite{vekaria2023inventory} (669); \cite{misinfo-dataset} (3K); & 660\\
Disinformation  & \cite{climate-disinfo-dataset} (11); \cite{moore2023consumption} (1.6K); \cite{fakenews-dataset} (2.1K) & \\ \hline
Typosquatting   & \cite{pouryousef2020extortion} (10.5K) & 14  \\ \hline
Phishing        & \cite{pouryousef2020extortion} (72.1K) & 309 \\ \hline
Piracy          & \cite{piracy-dataset} (2.7K)           & 136 \\ \hline
Sanctioned   &  \cite{ofac-list} (53)                  & 6   \\ 
\bottomrule
\end{tabular}}
\caption{Categories and counts of problematic publishers in each category used by our study (\cf \Cref{sec:methodology:pooling})}
\label{tab:problematic-dataset}
\vspace{-2mm}
\end{table}

\para{Discovering vulnerabilities with static analysis.}
We use static files of the standards (\ie \ads and \sellers) to detect dark pooling vulnerabilities. 

\parait{Curating a dataset of standards files.}
First, we crawl \ads files from the root of each problematic publishers we are studying, as well as from the Tranco Top-1M websites \cite{pochat2018tranco}. We extract the distinct domains of ad-networks listed in these \ads files (\ie domains associated with \ads DIRECT and RESELLER entries). 
Next, we crawl the corresponding \sellers files from the root of each previously identified seller domain. We extract the distinct ad-network domains listed in these \sellers files (\ie domains associated with \sellers INTERMEDIARY and BOTH entries). 
We repeat this process until no new seller domains/entities are discovered.
This ensures complete coverage of all supply chain entities involved in the sale of ad inventory on our problematic publishers and the Tranco top websites.

\parait{Identifying dark pools from the standards files.}
We use the above data to identify dark pools by finding all publishers whose \ads files share a common seller ID with another publisher on some ad-network. 
We use the \sellers file of the corresponding ad-network to determine the owner of such pooled seller IDs (\ie the {\em owner domain}). 
The ad-network that allows such pooling is referred to as the {\em pooled domain}.
To determine if a pool of publishers sharing a seller ID is a dark pool, we use DuckDuckGo’s entity list \cite{DDG-trackerradar} to map each publisher in the pool to its parent organization. 
To ensure 100\% completeness and accuracy, we manually validated organizational relationships of the small number entities that weren’t auto-matched in the DuckDuckGo’s list.
If a pool contains publishers owned by more than one parent organization and includes at least one of the problematic publishers listed in \Cref{tab:problematic-dataset}, then it is classified as dark pool.
In total, our static analysis yielded {26.1K} dark pools. These pools involved 399 problematic publishers, 1.7K unique victim publishers, and 962 unique ad-networks.

\para{Discovering vulnerabilities with dynamic analysis.}
As highlighted in prior work \cite{vekaria2023inventory}, static analysis alone cannot prove that dark pooling is actually occurring. 
This is because each publisher is responsible only for the content of their own \ads files and have no control over the misrepresentations in other publishers’ \ads files.
Therefore, we also perform dynamic analysis to gather evidence that a problematic publisher is actually monetizing its ad inventory using another publisher's seller ID. 

\parait{Identifying entities associated with dark pools from live page loads.}
We follow the methodology developed by Vekaria et al.~\cite{vekaria2023inventory}. {Following the crawling configuration and disclosures recommendations of Ahmad \etal \cite{ahmad2020apophanies}, we used a stateless web crawler driven by Selenium (v4.1.0) and the Chrome browser (v117.0) with bot mitigation strategies and Xvfb from a non-cloud vantage point to crawl problematic publishers and capture HTTP archive (HAR) files}. 
During each dynamic crawl, we load the problematic publisher, we wait 30 seconds for all resources, including ads, to finish loading. We click on the DOM elements associated with each ad URL and wait for the advertiser's website to be loaded. We log the associated URL, the chain of redirects, requests, responses, and payloads using the HAR file format.
Then, from our HAR files, we extract any strings which have a `key:value' or `key=value' format. We examine if any seller ID associated with our static dark pools appear in these extracted pairs.
If they do, we log the crawled website (publisher), the pooled seller ID, the ad-network that issued the pooled seller ID, the domain to which the pooled seller ID was issued to (owner domain), and the advertiser whose ad was served.
In a normal transaction, we expect that the owner domain matches the crawled publisher website, or that they are at least organizationally-related. However, in our current scenario examining dark pools containing problematic publishers, the owner domains are (victim) publishers who are not associated with the problematic publishers.
In total, our dynamic analysis yielded 1.3K dark pools which included 200 unique ad-networks, 347 unique owner domains, and 889 unique pooled seller IDs. The problematic publishers in these dark pools were observed with ads from 635 unique advertisers.

\para {Definition of vulnerability and its remediation.}
In our study, a ``vulnerability'' is a potential for unauthorized use or abuse of advertising ID(s) by malicious entities, and a vulnerability is considered ``fixed'' or ``remediated'' if we detect that the vulnerable advertising ID of a given entity is no longer abused by malicious entities. 
We consider any form of (partial or complete) reduction in vulnerabilities as remediation. This is because complete removal may not always be guaranteed due to multiple external factors such as business contracts in-place or internal decisions. These are beyond our scope/control and we acknowledge as a limitation (\Cref{sec:discussion:limitations})

\subsection{Designing multi-stakeholder notifications} \label{sec:methodology:notifications}

\para{Identifying notification recipients.}
We sent notifications to the three types of stakeholders (\cf \Cref{sec:background:supplychain}) in three separate rounds. This allowed us to accurately measure each stakeholder group’s influence in remedying the vulnerabilities.

\parait{Round 1: Notifying victim publishers.} We notified all popular (Tranco top-10K) publishers involved in dark pools with one of our problematic publishers, identified through our static analysis. We identified 1.7K such publishers.

\parait{Round 2: Notifying ad-networks.} We notified all of the ad-networks that facilitated {\em more than one} dark pool with one of our problematic publishers, identified through our static analysis. We identified 644 such ad-networks as recipients.

\parait{Round 3: Notifying advertisers.} We notified all the advertisers whose creatives were observed on problematic publishers due to dark pooling. We identified 635 unique advertisers.

\parait{Ordering of stakeholders.} This is a crucial design choice that multi-stakeholder notifications should evaluate as it can impact how different stakeholders respond and remediate vulnerabilities. 
In our case, the ordering of notifications was decided based on our intuition about which stakeholders would be the most influential at remedying vulnerabilities. 
We hypothesize that notifications to victim publishers are least influential because they are not monetarily impacted (at least directly) by the vulnerability and also do not have the ability to influence pool memberships directly.
Ad-networks, on the other hand, have the ability to re-assign pool memberships when they are the issuers of the pooled seller ID. 
However, doing so has the potential to lower the revenue generated by their pooled inventory.
In contrast, we hypothesize that notifications to advertisers will be the most influential since they are the most affected stakeholder, from monetary and reputational perspectives.
Ahmad \etal~\cite{ahmad2024companies} shows that brand-safety most directly impacts advertisers when their ad ends up on a problematic website. 
Advertisers face direct consumer backlash in the form of boycotts and negative press attention as a result of which they are incentivized to not advertise on such publishers, corroborating our hypothesis.
Additionally, their ability to control the flow of revenue to other stakeholders also makes them more capable of influencing their actions.
Therefore, our ordering (notifying publishers, then ad-networks, and finally advertisers) is expected to leave enough unresolved dark pools to allow for validity and statistical rigor during analysis in each round.

\parait{Design choice of parallel vs. sequential notifications} 
We adopt sequential multi-round approach for notifications as opposed to simultaneous or parallel notification approach due to a strong interconnectedness between different stakeholders. 
Typically publishers are listed on ``\textit{multiple}'' popular ad-exchanges and advertisers typically buy ads across ``\textit{multiple}'' popular ad-exchanges -- making it impossible to attribute the remediation to a specific notification, if notified simultaneously. 
For example, notifications to a subset of publishers could result in them reaching out to the respective ad-exchanges they work, who may also have been separately notified in the parallel notification strategy. 
In this case, it would be impossible to attribute remediation to a specific publisher or ad-exchange. We were unable to partition publishers, ad-exchanges, and advertisers into disjoint control and treatment sets so as to avoid such ``contamination'' in the parallel notification approach. 
However, in the sequential multi-round approach, our sequential notifications across different rounds focus on notifying unresolved vulnerabilities in each round, which allows us to easily attribute a remediation to the stakeholder in the latter round. 
The time gap of several weeks between different rounds mitigates the risk of contamination. 
There remains a non-zero risk but it is much smaller in sequential notifications than in parallel notifications. 
We further discuss this as part of the limitations (\cf \Cref{sec:discussion:limitations}).

\para{Establishing notification channels with recipients.}
We used emails to communicate with notification recipients, similar to the most prior works. 
We obtained email addresses from four sources outlined below and attached a tracking pixel to our notifications to see if recipients opened them.

\parait{Source 1: Contact pages.}
We searched the recipient’s homepage DOM for links containing the word “contact” to find the URL of the contact page. We then scraped these pages to retrieve email addresses using the regex: \verb|[\w\.-]+@[\w\.-]+\.\w+|. This method works only with English websites.

\parait{Source 2: \ads files.}
We extracted publisher email addresses from their \ads files, whenever available. A typical \ads email address is present in the following format: ``\texttt{CONTACT=adstxt@bbc.com}''. 
Per the \ads standard, this email address is meant to represent the point of contact to report issues regarding \ads file of the associated publisher domain.

\parait{Source 3: \sellers files.}
We extracted ad-network email addresses from their \sellers files. These addresses are extracted from the \texttt{contact$\_$email} key in the \sellers file.
Per the \sellers standard, this email address may be used to contact the Advertising System for questions or inquiries about the associated \sellers file.

\parait{Source 4: Common email prefixes.}
As a final resort, we also consider email prefixes commonly seen in contact emails by different companies. These included \texttt{info@}, \texttt{support@}, \texttt{help@}, \texttt{webmaster@}, and \texttt{contact@}. 
If the other methods did not provide an email address, we tested the availability of these common addresses. To ensure deliverability of our notifications, we sent test emails from an alternate account to monitor for bounces, preserving sender reputation of the main account used for the notifications.

\para{Varying sources of notifications.}
To measure the influence of the background and status of notification sources on the vulnerability resolution we randomly divide the notification recipients in each round into three equal-sized groups: two treatment groups and one control group. Entities were assigned to these groups at the start of each notification round.

\parait{Treatment 1: Academics as notification sources ($\mathcal{T}_1$).}
We used a dedicated university email address for notifications to recipients in this group. 
These notifications had university branding and mentioned our university affiliations.

\parait{Treatment 2: Activists as notification sources ($\mathcal{T}_2$).}
We collaborated with the Check My Ads (CMA) institute \cite{checkmyads} (previously known as ``Sleeping Giants'' \cite{li2021beyond, braun2019activism}), a well-known activist organization which names-and-shames ad-tech entities that aid the monetization of problematic content.
Unfortunately, due to organizational limitations (\cf \Cref{sec:discussion:limitations}), we couldn't use CMA's servers for sending these emails and used our university servers instead.
However, recipients in this group received notifications which had CMA branding and explicitly mentioned CMA's involvement.

\parait{Control group ($\mathcal{C}$).}
The control group received no notifications from us. 
They were used as a baseline from which the effects of $\mathcal{T}_1$ and $\mathcal{T}_2$ notifications could be measured.
Entities in this group were informed of their dark pooling vulnerabilities at the end of our 9-month long study.

\para{Curating notification content for stakeholders.}
All notification emails contained brief introductions to the researchers along with a link to our project webpage. 

\parait{The project webpage.}
The webpage (\textit{\url{ad-inventory-fraud.github.io}}) contained a description of our research project's goals and dark pool identification methodology.
This page also contained FAQ and Contact sections to help recipients better understand their notifications or reach us. 
In our descriptions and notifications, we consciously avoid alarmist or accusatory language. For example, we use the term `ad safety research' in lieu of `ad fraud research' and consistently describe the identified vulnerabilities as `potential vulnerabilities' and problematic publishers as `potentially unsafe websites'. 
This is done to develop a positive relationship with the entities and show intent to aid vulnerability resolution.

\parait{Subject lines of notification emails.}
Notification emails to advertisers had the subject line: {\em Brand safety violation for your domain {\tt <advertiser.com>}}.
Other notification emails to ad-networks and publishers had the subject line: {\em Potential ad inventory vulnerability for your domain {\tt <domain.com>}}.

\parait{Body of notification emails.}
In addition to an introductory preamble described earlier, the email body contained summaries of the vulnerabilities identified by our research. We crafted unique summaries for each stakeholder.

    \noindent \ding{182} {\em Victim publishers.} We report vulnerabilities (\ie from static and dynamic analysis) using the following text summaries:
    (1) For vulnerabilities detected via dynamic analysis: {``During network traffic analysis of problematic publishers, we observed that {\tt <num>} sellerID(s) in your \ads are being used by {\tt <num>} potentially problematic publisher(s) to monetize their ad inventory.''}; and
    (2) For vulnerabilties detected via static analysis: {``We observed that {\tt <num>} sellerID(s) in your \ads are being used in \ads of at least {\tt <num>} other potentially problematic publishers.''}
    
    \noindent \ding{183} {\em Ad-networks.} We report two types of vulnerabilities from each of our analysis approaches using the following text summaries:
    (1) For vulnerabilities detected by static analysis of \ads files: {``{\tt <num>} seller IDs issued by you are being pooled by {\tt <num>} potentially problematic publishers.''};
    (2) For vulnerabilities detected by static analysis of \sellers files: {``{\tt <num>} seller IDs owned by you and issued by another ad-network are being pooled by {\tt <num>} potentially problematic publishers.''};
    (3) For \ads vulnerabilities confirmed by dynamic analysis: {``We confirmed that {\tt <num>} seller IDs issued by you are being pooled by {\tt <num>} potentially problematic publishers.''}; and 
    (4) For \sellers vulnerabilities confirmed by dynamic analysis: {``We confirmed that {\tt <num>} seller IDs owned by you and issued by another ad-network are being pooled by {\tt <num>} potentially problematic publishers''}.
    
    \noindent \ding{184} {\em Advertisers.} Because we do not expect advertisers to be particularly technically adept, we frame our notification around the presence of their ad creatives on potentially problematic content as follows: {``An ad creative associated with your brand was observed on {\tt <num>} problematic publishers. This association could negatively impact your reputation and future business.''}

\parait{Reports attached to notification emails.}
To provide concrete evidence to support the reported vulnerabilities, we generated an automated PDF report, with appropriate branding based on the notification source. 
This report contained: (1) up to 25 instances of vulnerabilities which involve the specific entity; (2) an explanation of the root causes and implications of each vulnerability; and (3) possible remediation options.
In addition, notifications to advertisers also included attachments of screenshots showing the advertiser's ad creative on a problematic publisher and the HAR file associated with the page load.
Overall, similar notification framing was used for each stakeholder.

\begin{figure*}[t]
    \centering 
    \includegraphics[width=\linewidth]{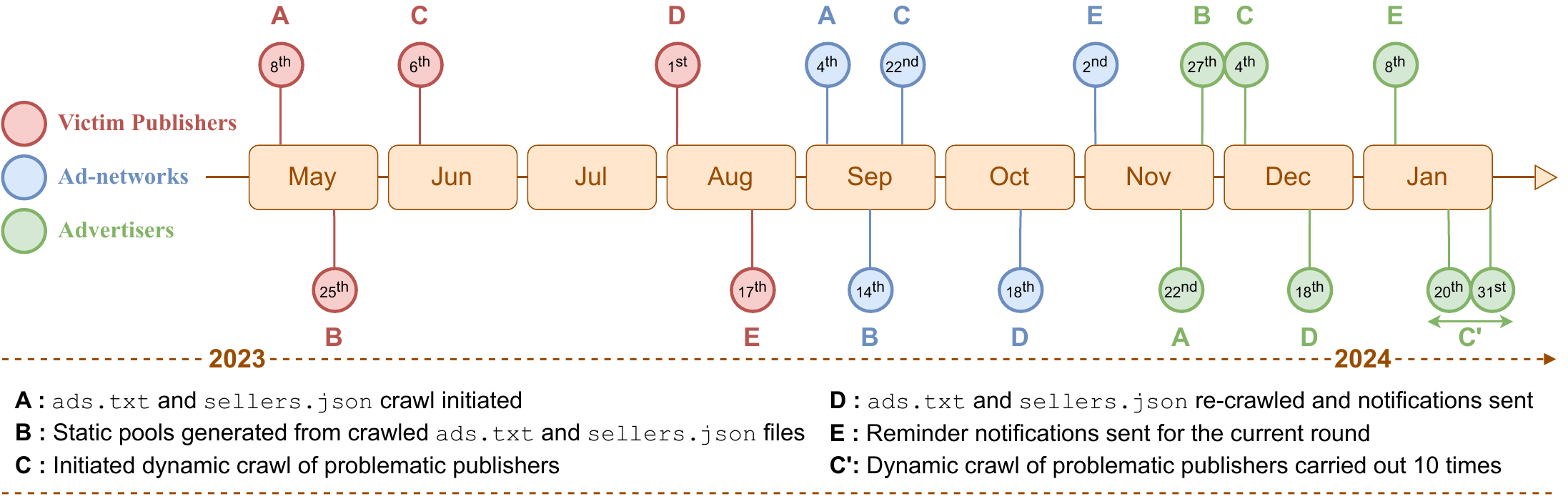}
    \caption{An overview of the timeline of the the notification campaign.}
    \label{fig:timeline}
\end{figure*}

\para{Timeline of notifications.}
\Cref{fig:timeline} depicts the timeline of our notification campaign.
Prior to each round of notifications we conduct static and dynamic analysis to identify dark pooling vulnerabilities w.r.t to our dataset of problematic publishers (\cf \Cref{sec:methodology:pooling}).
Then, we generate and send notifications and reports for each entity associated with that round. 
If we do not hear back, we send a reminder email approximately 1-2 weeks after the initial email. We then wait a month before re-conducting our static and dynamic analysis.
We use the data from this analysis to assess the influence of notifications sent during the current round. 


\section{Recipient Engagement with Notifications} \label{sec:engagement}

Our study included a total of 2.9K unique entities involved in or impacted by dark pooling in the ad-tech supply chain. 
These comprise of 1.7K victim publishers, 644 ad-networks, and 635 advertisers.
In this section, we report the rates (\Cref{sec:engagement:rates}) and the nature (\Cref{sec:engagement:nature}) of notification engagement.

\subsection{Rate of engagement with notifications} \label{sec:engagement:rates}

\para{Notification delivery rates.}
Of the 2.9K entities involved in our study, we only sent notifications to two-thirds (\cf \Cref{sec:methodology:notifications}) and used the remaining one-third as a control group against which the effect of our notifications could be determined.
Of these 1.9K notifications (across both treatment groups $\mathcal{T}_1$ and $\mathcal{T}_2$), 79.5\% were successfully delivered and opened by their recipients. This rate of delivery exceeds the rates reported in prior notification research \cite{cetin2017make,cetin2016understanding,stock2016hey, stock2018didn}.
We hypothesize that this is because of our ability to leverage email addresses from \ads and \sellers files for contacting publishers and ad-networks, respectively.
The remaining notifications (417 notifications or 20.5\%) failed to be delivered for a variety of reasons including emails rejected by spam filters (40.2\% of failures) and invalid email address ({21.5\%} of failures).

\para{Notification response rates.}
Of the 1.5K recipients who read our notifications, we received an email response from 215 (14\%). Of these, 147 emails were generic responses that appeared to be template responses. The remaining 68 emails were engaging directly with the content of our notifications and were qualitatively analyzed (\cf \Cref{sec:engagement:nature}). 
Of these, 40 were from victim publishers (3.5\% response rate), 18 were from ad-networks (4.2\% response rate), and 10 were from advertisers (2.9\% response rate).
This is in line with the response rates observed in prior research \cite{maass2021snail, maass2021effective, stock2018didn, stover2023website, soussi2020feasibility, utz2023comparing} focused on single-stakeholder notifications. It should be noted that the absence of an email response does not suggest that the notifications did not influence vulnerability resolution. We measure the exact influence on vulnerability resolution in \Cref{sec:impact}.

\subsection{Nature of engagement with notifications} \label{sec:engagement:nature}

\para{Methodology for thematic analysis.}
To systematically explore different themes emerging in responses to our notifications, we adopted the state-of-the-art thematic analysis methodology \cite{braun2006using, stover2023website} to identify, analyze, and report patterns in the responses. This analysis can uncover common attitudes towards and challenges faced in the ad-tech vulnerability resolution. 
This approach eliminates the need for multiple coders, so one author familiarized themselves with the email responses by reading them multiple times. Next, a codebook was developed by the same author by assigning codes to each sentence in the response emails. This codebook was iteratively refined, in consultation with the other authors, to consolidate the final set of codes in our codebook. 
This is because \cite{clarke2013successful} emphasizes that thematic analysis is reflexive and interpretive rather than a reliability exercise, so one researcher can reasonably undertake initial coding to ensure immersion and consistency, with credibility and analytic quality ensured through subsequent collaborative refinement and consensus, eliminating the need for statistical reliability.
Next, the authors use this codebook to independently code all the emails. 
We follow Stover et. al~\cite{stover2023website}, who adopt Clarke and Braun’s~\cite{clarke2013successful} understanding of thematic analysis without requiring to calculate inter-rater reliability.
Instead, similar to their work, the coder convened with the second author to deliberate any ambiguities and reach a consensus on the final coding in order to ensure a high quality of analysis as suggested in the approach. 
Finally, the email responses were grouped into three high level themes: (1) effort towards resolution, (2) demonstration of concern or awareness, and (3) lack of trust or resources. 
The codebook, identified themes, categories, and sub-categories are presented in Table~\ref{tab:codebook}. 
Next, we use the developed codebook categories to study responses by themes, recipient type, and sender reputation (\emph{Cf.} \Cref{tab:thematic-responses} (appendix)).

\begin{table*}[t] 
\caption{Code book with themes, categories, and subcategories from the thematic analysis of notification responses.}
\label{tab:codebook}
\resizebox{\linewidth}{!}{%
\begin{tabular}{p{0.24\linewidth} p{0.8\textwidth} >{\centering\arraybackslash}p{0.07\textwidth}}
\hline
\\
\textbf{Category} & \textbf{Sub-categories} & \textbf{Frequency} \\
\\
\hline
\multicolumn{3}{l}{} \\
\multicolumn{3}{l}{\multirow{-2}{*}{\textbf{Theme 1: Effort towards resolution}}} \\
\hline
\\
\scalebox{0.9}{Actionable details requested} &
\scalebox{0.9}{%
\begin{tabular}[t]{@{}p{0.9\textwidth}@{}}
asked for additional details (13); seeking clarification on conveyed information (11); asked technical questions (4); asked methodology to contact problematic entities (2); data collection time period details requested (2); requested the full report (2); tried to identify root cause of the issue (2); verifying if vulnerability is due to reseller relationships (2); action items requested (1); actionable information acknowledged (1); ad screenshot requested to see evidence of issue (1); confirming the external inventory scanning tool used by us (1); discussed possible solution to implement (1); Seeking details on how to fix the issue (1); technical details shared (1)
\end{tabular}} &
\scalebox{0.9}{28} \\
\\
\scalebox{0.9}{Forwarded to the concerned team} &
\scalebox{0.9}{%
\begin{tabular}[t]{@{}p{0.9\textwidth}@{}}
shared with the relevant department (7); will share with the relevant department (7); security reporting procedure explained (5); asked to contact other department (3); escalated the issue (2)
\end{tabular}} &
\scalebox{0.9}{22} \\
\\
\scalebox{0.9}{Collaborative resolution performed} &
\scalebox{0.9}{%
\begin{tabular}[t]{@{}p{0.9\textwidth}@{}}
scheduled a meeting (10); suggested collaborative issue resolution (3); engaged in phone call (1); suggested collaborative business promotion (1)
\end{tabular}} &
\scalebox{0.9}{14} \\
\\
\scalebox{0.9}{Fixed the reported issues} &
\scalebox{0.9}{%
\begin{tabular}[t]{@{}p{0.9\textwidth}@{}}
made fixes (7); asked to verify current status of issues post resolution (2); investigated the issues (2); removed off the vulnerable entries present in the report (2); actively fixing the issue (1); ads.txt added but never used (1); sellers.json had stale entries (1); will not engage in further correspondence on security issues (1)
\end{tabular}} &
\scalebox{0.9}{11} \\
\\
\scalebox{0.9}{Contacted responsible entities for fix} &
\scalebox{0.9}{%
\begin{tabular}[t]{@{}p{0.9\textwidth}@{}}
will reach out to responsible entities (6); reached out to responsible entities (3)
\end{tabular}} &
\scalebox{0.9}{9} \\
\\
\scalebox{0.9}{Justifying the reported issue} &
\scalebox{0.9}{%
\begin{tabular}[t]{@{}p{0.9\textwidth}@{}}
defending few reseller relations (1); justifying low inventory fraud rate due to the usage of brand-safety tools (1); questioning vulnerabilities as benign due to direct relations (1); stating to work via mostly direct-business relations (1)
\end{tabular}} &
\scalebox{0.9}{4} \\
\\
\hline
\multicolumn{3}{l}{} \\
\multicolumn{3}{l}{\multirow{-2}{*}{\textbf{Theme 2: Demonstration of concern or awareness.}}} \\
\hline
\\
\scalebox{0.9}{Acknowledged the notification} &
\scalebox{0.9}{%
\begin{tabular}[t]{@{}p{0.9\textwidth}@{}}
acknowledged the notification positively (47); grateful for sharing the information (2)
\end{tabular}} &
\scalebox{0.9}{47} \\
\\
\scalebox{0.9}{Motivated to fix the issue} &
\scalebox{0.9}{%
\begin{tabular}[t]{@{}p{0.9\textwidth}@{}}
will investigate on the issue (13); will resolve the issue (8); willing to help fix the issue (3); will remove ads.txt (1); will remove vulnerable entries from sellers.json (1)
\end{tabular}} &
\scalebox{0.9}{21} \\
\\
\scalebox{0.9}{Concerned about the notification} &
\scalebox{0.9}{%
\begin{tabular}[t]{@{}p{0.9\textwidth}@{}}
concerned about the issue (5); confused in interpreting intended versus actual recipient (3); activists are anti-Fox where Fox is our biggest customer (1); cared about phrasing/wording of the notification (1); concerned about the already incurred effect on reputation (1); concerned about the potential loss of ad revenue already incurred (1); contacted consultant about the notification (1); curious about recipient selection for notification (1); demonstrated seriousness about the notification (1); expressed surprise from the details in the notification (1); lawyer reached out (1); misinterpreted reaching out to hosting provider (1); notification perceived as frustratingly alarmist in nature (1); were afraid of the effect of the notified issue on business revenue (1); worried about vulnerability's impact on web traffic (1)
\end{tabular}} &
\scalebox{0.9}{15} \\
\\
\scalebox{0.9}{Research initiative appreciated} &
\scalebox{0.9}{%
\begin{tabular}[t]{@{}p{0.9\textwidth}@{}}
research perceived as interesting (3); received appreciation for research (2); admired the research (1); considered initiative to be similar to another initiative (1); considered work being done as very important (1); conveyed appreciation for initiative (1); expressed willingness to connect in the future (1); the initiative perceived as interesting (1); showed support for initiative (1)
\end{tabular}} &
\scalebox{0.9}{9} \\
\\
\scalebox{0.9}{Awareness about the vulnerability} &
\scalebox{0.9}{%
\begin{tabular}[t]{@{}p{0.9\textwidth}@{}}
were already aware of the issue (2); advocated awareness around the issue in the industry (1); were aware of our research via publisher notification (1); were unaware of the issue (1)
\end{tabular}} &
\scalebox{0.9}{4} \\
\\
\scalebox{0.9}{Notification unhelpful or uninterested} &
\scalebox{0.9}{%
\begin{tabular}[t]{@{}p{0.9\textwidth}@{}}
considered notification as not helpful (1); demonstrated unseriousness about the notification (1); no documentation (1); not interested in the notification (1); perceived no actionable information (1)
\end{tabular}} &
\scalebox{0.9}{4} \\
\\
\hline
\multicolumn{3}{l}{} \\
\multicolumn{3}{l}{\multirow{-2}{*}{\textbf{Theme 3: Lack of trust or resources.}}} \\
\hline
\\
\scalebox{0.9}{Doubting correctness of the report} &
\scalebox{0.9}{%
\begin{tabular}[t]{@{}p{0.9\textwidth}@{}}
asked for the definition of problematic (7); felt that the report had some inaccuracies (2); doubted the association of reputable big ad-networks with the vulnerability (1); doubted the data viability (1); felt that the report didn't make sense (1); not sure if the report was accurate (1); notified about 0 issues (1); verified no threats to be present (1)
\end{tabular}} &
\scalebox{0.9}{13} \\
\\
\scalebox{0.9}{Misinterpreted as phishing} &
\scalebox{0.9}{%
\begin{tabular}[t]{@{}p{0.9\textwidth}@{}}
authentication of the sender (6); misinterpreted as a phishing attempt (2); company policy prevented opening external documents (1); considered notifications as spam (1)
\end{tabular}} &
\scalebox{0.9}{9} \\
\\
\scalebox{0.9}{Limited resources to fully resolve} &
\scalebox{0.9}{%
\begin{tabular}[t]{@{}p{0.9\textwidth}@{}}
expressed to be a small publisher (2); considered time-consuming to reach out to publishers (1); expressed a lack of resources to fix the issue (1); unsure how to handle issues with a large number of domains (1)
\end{tabular}} &
\scalebox{0.9}{4} \\ 
\\
\scalebox{0.9}{Complete resolution beyond control} &
\scalebox{0.9}{%
\begin{tabular}[t]{@{}p{0.9\textwidth}@{}}
expressed no say in the removal of their entry from other ads.txt (1); expressed zero control over reseller accounts (1); removal from other ads.txt not in our control (1)
\end{tabular}} &
\scalebox{0.9}{3} \\
\\
\hline
\end{tabular}%
}
\end{table*}
\para{Response theme 1: Effort towards resolution.}
We found that \vpubs and \adxs often requested additional actionable information or clarifications to aid the resolution of our reported issues. This is not commonly seen in responses from \brands, however.
We hypothesize that this is because \brands are less likely to maintain in-house tech-teams and instead rely on ad agencies to manage their ad campaigns for them. This hypothesis is supported by the high number of \textit{forwarded to the right department} category responses observed from \brands.
This suggests that some of the generic contact emails from an advertiser's website may not be particularly useful for performing \brand notifications related to ad-fraud. 
In contrast, \vpubs yielded a mix of responses under this category -- smaller publishers did not have separate teams for advertising, however, larger publishers managed advertising under a separate team to which our notifications were forwarded.
We found that many of the responsive \vpubs and \adxs were motivated to \textit{collaboratively resolve the issues} and scheduled either a video meeting or a phone call to understand and resolve the reported vulnerability. 
Victim publishers were most likely to have responded following a successful resolution of the reported vulnerability. For example, one \vpub responded -- \quoteResponse{We've cleaned up our \ads to only active providers. Can you check it again?} -- suggesting that they do not regularly update their \ads files to maintain them up-to-date. Untimely updates can aid problematic publishers to monetize through victim publisher's sellerIDs.

\para{Response theme 2: Demonstration of concern or awareness.}
Responses to our notifications were largely positive, with many including a note of thanks for our report and showing concern and motivation to resolve the vulnerability. 
Even those that started as negative eventually turned positive as engagement continued. For example, one entity who initially responded \quoteResponse{I find your email frustratingly alarmist and devoid of actionable information} eventually appreciated our notifications and research after several back-and-forth emails in which we provided clarifications and help towards vulnerability resolution. 
Some publisher recipients were eager to act but didn't know how to go about resolving dark pools. For example, \quoteResponse{We truly want to prevent dark pooling from happening. What do you suggest that we should do? \dots can we reach out to [ad-networks] and ask them to monitor and provide logs for the usage of vulnerable sellerIDs that you provided?}
Many entities showed concern regarding the indirect effects of the reported vulnerabilities on their reputation and revenue. For example, one entity responded: \quoteResponse{is there some way of estimating how much potential harm may have been caused to our website — as far as reputation and ad revenue go — from the kinds of activity that you identified in your research?}, later adding \quoteResponse{if these ad inventory vulnerabilities have played a part in [our] revenue decline, I'd really like to know to what extent they have — and what we can do to get back any potential loss in revenue}. 
Other responses showed awareness about the types of vulnerabilities reported and appreciation for our research efforts. 
For example, one entity responded -- \quoteResponse{This matter is near and dear to me, and I've spoken about it publicly at [ad-tech conference] as well as on the [podcast]} ... \quoteResponse{Beyond all this, I think the work you're doing is necessary for our industry.}

\para{Response theme 3: Lack of trust or resources.}
Several entities highlighted their inability or unwillingness to resolve the reported vulnerabilities for a variety of reasons ranging from doubting the correctness of the report to lack of technical resources to properly understand and resolve the vulnerabilities.
In one interesting case, an \adx claimed our report (which contained evidence that a misinformation publisher was pooling a seller ID issued by them) was inaccurate --- \quoteResponse{My tech team verified that all the references on your list are legitimate publishers owned by one of the corporate entities on our small inclusion list}; yet, within hours, this same \adx updated their \sellers file with a comment \quoteResponse{We no longer work with [misinformation publisher] or associated properties.}
Several entities also suspected our notifications to be part of a phishing campaign. 
These entities often sought proof of our identities, with some even reaching out to our university's media relations team for clarification.
\quoteResponse{... a lawyer who received an email from your address yesterday is concerned about the issue that this email may have been a phishing attempt.}. Fortunately, these suspicions were eventually resolved favorably.
Finally, many publishers reported their inability to resolve the reported vulnerabilities due to the lack of technical resources --- for example, one publisher reported \quoteResponse{I’ve made some immediate fixes — basically, just cleaning up our \ads file, removing the lines in question but also deleting a lot of probably unnecessary other lines as well. Beyond that, I’m not sure how to contact all the various other players — the ad-networks, the problematic websites. We’re a very a small publisher...}. 
This lack of technical know-how was also found to be the cause of several reported vulnerabilities, as one publisher reported -- \quoteResponse{I had no idea that there was any issue with our \ads file. I’ve always just included whatever lines that [AdX] tell us to add.}

Overall, we observe that notifications sent as \academics received less responses as compared to \cma for \vpubs. However, \adxs show more positive responses towards \academics than \cma. This is a very interesting insight. CheckMyAds (i.e., \textit{activist}) often publicly calls out \adxs on social media platforms shaming them for monetizing problematic outlets. This seems to have lost the trust of \adxs. 
In line with this finding, one of the \adxs mentioned -- \quoteResponse{... admired the research, but your site shows a lot of anti-Fox News posts, and they're one of our biggest publishers}.

\section{Notification Impact on Dark Pooling} \label{sec:impact}

Our qualitative analysis on responses to notifications does not provide a complete picture on their impact because recipients may initiate remedial actions without responding to the notification. 
In this section, we examine 3 measures for assessing the influence of notifications (\Cref{sec:impact:assessing}). We then use these to evaluate how the notification source (\Cref{sec:impact:sender}) and recipient (\Cref{sec:impact:recipient}) influence remedial responses.

\subsection{Assessing the impact of notifications} \label{sec:impact:assessing}

\begin{figure}
    \centering
    \includegraphics[width=1\linewidth]{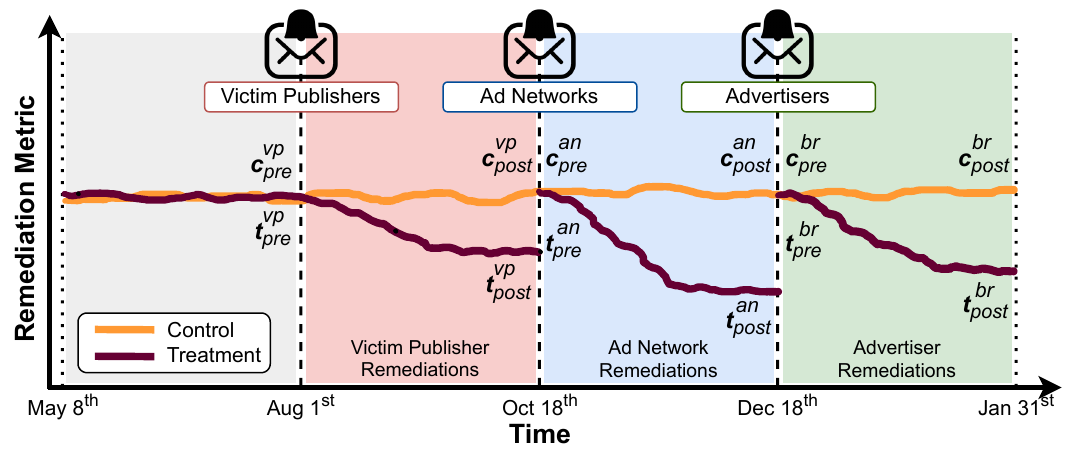}
    \caption{An illustration of the measured remediation measures before and after notifications to different stakeholders in different rounds. The difference-in-differences is computed from these measures as $\Delta_{(t,c)} = (t_{post} - c_{post}) - (t_{pre} - c_{pre})$ for each ad-tech entity.}
    \label{fig:pm-did}
\end{figure}

\para{Propensity matched difference-in-differences (PM-DiD).}
Our goal is to uncover the overall impact of notifications on specific measures of dark pooling relevant to each stakeholder. 
We use {\em propensity-matched difference-in-differences}, a common statistical approach for identifying the causal effect size of an intervention when a randomized controlled trial is not possible \cite{callaway2021difference, donald2007inference, tanveer2023glowing}.
We randomly divide the notification recipients in each round into three equal-sized groups: two treatments and one control.
We apply PM-DiD in 3 steps as shown in Algorithm 1 (\Cref{appendix:PMDiD}).

\parait{Step 1: Matching entities in treatment and control groups.}
First, we create a control group that is statistically similar to the treatment group based on specific dark pooling measures.
Our treatment group may be $\mathcal{T}_1$, $\mathcal{T}_2$, or $ \mathcal{T}_1 \bigcup \mathcal{T}_2$ and our control group is some subset of $\mathcal{C}$ (\cf \Cref{sec:methodology:notifications}).
For each entity in our treatment group, and for each comparison measure, we find its pre-notification nearest neighbor---the entity in the control group candidates which is the most similar for the given measure (\ie has the value of that measure numerically most close to the treatment entity in consideration) and include this candidate in our control group. 
This helps us to establish a baseline where the pre-notification differences between treatment and selected control are minimal and the effect observed post notification can be attributed to the performed treatment.

\parait{Step 2: Computing pre- and post-intervention measures.}
Next, we identify the pre- and post-intervention measures for each entity in the treatment group ($t$) and their matched control counterpart ($c$). We record pre-intervention measure ($t_{pre}$ and $c_{pre}$) from crawls that occur 1-2 weeks before sending notifications to the treatment entity. We record post-intervention measures ($t_{post}$ and $c_{post}$) from crawls that occur 4-5 weeks after sending notifications. \Cref{fig:pm-did} shows an illustration of effective remediations.

\parait{Step 3: Computing effect sizes and statistical significance.}
We compute the effect size as: $\Delta_{(t,c)} = (t_{post} - c_{post}) - (t_{pre} - c_{pre})$. A negative value denotes that the corresponding measure is lower post-treatment. 
We repeat this to compute $\Delta$ for each matched $(t,c)$ pair. We then report: (1) $N_{\text{rem.}}$: the number of $(t, c)$ pairs that showed a remedial action (\ie $\Delta_{(t,c)} < 0$); (2) $\mu_{\text{ov.}}$ the mean effect size across all $(t, c)$ pairs; and (3) $\mu_{\text{rem.}}$: the mean effect size across all $(t,c)$ pairs that showed a remedial action.

\para{Measures of dark pooling for each stakeholder.}
We track the following dark pooling measures to evaluate the impact of our notifications on each stakeholder.

\parait{Victim publishers.} We focus on one key measure: the number of problematic publishers which pool and use the seller ID of the victim publishers (we refer to this count as \probpools for a publisher $vp$).
We expect \probpools to reduce if $vp$ engages with their ad-networks to either get: (1) an unpooled seller ID, (2) assigned a different pool than the problematic publishers, or (3) remove problematic publishers from their pool. 

\parait{Ad-networks.} We use two measures for ad-networks: (1) the number of seller IDs they issue that are used for dark pools (we refer to this as \darkpools for the ad-network $adx$); and (2) the number of seller IDs issued by other ad-networks to $adx$ that facilitate dark pooling (we refer to this as \partnerdarkpools for $adx$).
We expect \darkpools to decrease if $adx$ corrects its own pool assignments to prevent pooling of problematic and popular domains.
We expect \partnerdarkpools to decrease if $adx$ engages with other ad-networks where it is listed to correct their facilitation of dark pools.

\parait{Advertisers.} Our notifications to advertisers only informs them about the ad-network responsible for displaying their ads on problematic publishers and the associated seller ID. Therefore, we expect a reduction in \darkpools for the responsible $adx$ if the advertisers request to remove such \darkpools. The fact that advertisers monetize the ad-tech supply chain would motivate ad-networks to act.

\subsection{Effect of notification recipient} \label{sec:impact:recipient}

\begin{table}[t!]
\caption{Effect of notification recipient on dark pooling.}
\centering
 \begin{tabular}{l c c c} 
 \toprule
 \textbf{Recipient (Measure)} & \textbf{$N_{\text{rem.}}$} & \textbf{$\mu_{\text{ov.}}$} & \textbf{$\mu_{\text{rem.}}$} \\
 \midrule
 Publishers (\probpools) & 54.0\% & 2.9 & -30.2 \\
 Ad-networks (\darkpools) & 81.6\% & -0.7 & -1.5 \\
 Ad-networks (\partnerdarkpools) & 76.9\% & 3.2 & -1.4 \\
 Advertisers (\darkpools) & 72.6\% & 0.7 & -0.7 \\
 \bottomrule
 \end{tabular}
 \label{tab:impact:recipient}
\end{table}

\Cref{tab:impact:recipient} shows the impact of notifications on dark pooling measures for each stakeholder. The data reveals several interesting findings. Notably, all stakeholders are generally responsive to our notifications with between 54-82\% of them showing a remedial action (\ie $\Delta_{(t, c)} < 0$).

\para{Ad-networks are most responsive; publishers are least responsive.}
At the less responsive end, 54\% of notified publishers took successful remedial action, reducing their participation in dark pools more than their control counterparts. In contrast, 81.6\% of notified ad-networks took successful remedial action, reducing the number of seller IDs used for dark pooling at a higher rate than their control counterparts. 
Rather surprisingly, despite controlling the revenue of the advertising supply chain, only 72\% of advertisers showed some mitigation in dark pools.
This finding might be explained by the fact that, among all the stakeholders, ad-networks can mitigate dark pools most easily because they control the assignment of seller IDs. 
Ad-exchanges are the central entity in the ad-tech supply chain as they work with both sellers (i.e., publishers to auction their ad inventory) and buyers (i.e., advertisers to bid and buy relevant ad inventory from). Since dark pools are a result of shared usage of these seller IDs assigned by the ad-exchanges to different sellers, they can be reasoned to be the most responsive.
On the other hand, publishers and advertisers need to engage with their ad-networks. Our qualitative analysis supports this, showing that many publishers were unsure how to resolve their participation in dark pools. 
Moreover, this analysis shows that ad-networks remediate dark pools with a higher probability when advertisers reach out to them as compared to when publishers do so -- likely because advertisers are the source of funding in the ad supply chain.

\para{Ad-networks are able to successfully engage with their counterparts to perform remedial actions.}
Our analysis also shows that ad-networks successfully engage with their counterparts to remediate dark pools that they belong to in 77\% of cases, as evident by the \partnerdarkpools measures. 

\para{The effect size associated with successful publisher remediation is the highest.}
Finally, the effect size associated with successful publisher remediation is significantly higher than other measures (\probpools $\mu_{\text{rem.}}$ = -30.2). This is because victim publishers are often in very large pools which have many problematic publishers. Therefore, a single remediation action, like assigning the publisher a new seller ID, drastically impacts $\mu_{\text{rem.}}$.

\subsection{Effect of notification source} \label{sec:impact:sender}

\Cref{tab:impact:sender} shows the impact of notifications sent by academics and activists on the stakeholders' remedial actions. The data indicates that, with one exception, there is no significant difference in remediation rates between notifications sent as academics and those sent as activists.

\para{Publisher responses are influenced by the notification source.} We compared the distributions of $\Delta_{(c,t)}$ for academic notifications ($t\in \mathcal{T}_1$) and activist notifications ($t\in \mathcal{T}_2$) using a $t$-test to see if stakeholder responses varied significantly by notification sources. We found that only publishers showed a statistically significant difference in their responses. Publishers were more likely to perform remedial actions when they received notifications from activists rather than from academics. This difference is evident in both the fraction of publishers that performed a remedial action and the mean effect size of the remedial action. This is likely due to CMA's long-standing reputation of publicly name-and-shame ad-tech entities by calling them out on social media for monetizing problematic content online. This is aligned with our findings in the qualitative analysis (\cf \ref{sec:engagement:nature}).

\begin{table}[t!]
\caption{Effect of notification source on measures of dark pooling for each stakeholder. $^*$ denotes a statistically significant ($t$-test; $p<$.05) difference in the overall effect based on notification source.}
\centering
 \begin{tabular}{c l c c c} 
 \toprule
 \textbf{Source} & \textbf{Recipient (Measure)} & \textbf{$N_{\text{rem.}}$} & \textbf{$\mu_{\text{ov.}}$} & \textbf{$\mu_{\text{rem.}}$} \\
 \midrule
 \multirow{5}{4em}{Academic} 
 & Publishers (\probpools) $^*$ & 52.3\% & 5.8 & -27.6 \\
 & Ad-networks (\darkpools) & 81.2\% & 0.8 & -1.7 \\
 & Ad-networks (\partnerdarkpools) & 75.3\% & 5.6 & -0.7 \\
 & Advertisers (\darkpools) & 72.6\% & 1.1 & -0.9 \\
 \midrule
 \multirow{5}{4em}{Activist} 
 & Publishers (\probpools) $^{*}$ & 55.6\% & -0.1 & -32.5 \\
 & Ad-networks (\darkpools) & 82.0\% & -0.6 & -1.3 \\
 & Ad-networks (\partnerdarkpools) & 78.4\% & 0.9 & -2.0 \\
 & Advertisers (\darkpools) & 72.6\% & 0.4 & -0.6 \\
 \bottomrule
 \end{tabular}
 \label{tab:impact:sender}
\end{table}

\section{Discussion and Conclusion} \label{sec:discussion}

\subsection{Validity of study design} \label{sec:discussion:validity}

\para{Internal and external validity considerations.}
Internal validity measures how well a study is conducted and how accurately its results reflect the studied group, while external validity assesses the generalizability of the findings. We evaluate different internal and external validity threats applicable to our study and explain how we handled them to minimize any validity issues. We follow recommendations from prior works \cite{campbell2015experimental, anderson2018design} to ensure the validity of our study and results.

\para{Threats from lack of representativeness.}
We ensured representativeness by notifying all entities participating in dark pooling, without imposing selection criteria for ad-networks and advertisers. But, problematic publishers often impersonate popular publishers. So, we selected our sample from the top-10K publishers whose advertising identifiers were used by problematic publishers in their \ads files. We randomly assigned the selected entities into different treatment and control groups to avoid selection bias. We applied identical treatment steps to different entities, though the specific vulnerabilities disclosed and remediation steps suggested varied based on their role in the ad-tech supply chain. We did not introduce any other experimenter bias.

\para{Threats from subject attrition.}
We did not measure the long-term effects of notifications on entities, so attrition concerns are not relevant. Additionally, ad-tech entities do not change roles over time (\eg a publisher does not become an ad-network). We ensured blinding by not informing participants about the applied treatment or intervention. We did not perform any experimental manipulation, such as entity-specific priming or incentives.

\para{Threats from situational factors.}
Notified entities are globally spread. We did not account for situational factors like the time and date of notification or location, which could affect the external validity of our results. It's a limitation.

\para{Threats from lack of trust and realism.}
To improve external validity, we aimed for psychological realism \cite{brewer2000research} by communicating through our email text and website content that we were working in the entity’s favor, helping to reduce their participation in a problematic activity. 
This approach was intended to develop their trust in us, increasing confidence in our notifications. Our notifications were also based on verifiable real data which reflected their true participation in dark pools, further bolstering realism and building trust.

\subsection{Limitations} \label{sec:discussion:limitations}

\para{Multi-stakeholder interdependence.}
The online advertising multi-stakeholder supply chain is fundamentally fragmented -- interdependent with mixed incentives for each participant.
Publishers and advertisers cannot directly fix their participation in dark pooling without cooperation from ad-networks. 
Ad-networks can address their own facilitation of dark pools but cannot control the behavior of other ad-networks using their platform.
Using a single round notification (notifying all stakeholders simultaneously) would make it impossible to identify which entity resolved the vulnerability. 
To solve this, we used a multi-round notification design, notifying one stakeholder in each round. 
This allows us to identify more confidently which stakeholder resolved the vulnerability. 
However, there is still a risk of contamination when a stakeholder responds much later, after the next round has begun. 
We minimized such risks by leaving 4-5 weeks between each round of notifications, giving each stakeholder ample time to respond and remediate.
Based on the email responses, we observed that nearly all arrived within two weeks of the initial notification.
Additionally, publishers may reach out to ad-networks to remediate the vulnerability, before we notify ad-networks in the second phase, alerting them. However, publishers are incentivized to work with more ad-networks and vice versa to maximize their respective revenue. Hence, we acknowledge that complete isolation is not possible.
However, since we detect vulnerabilities before each phase, ad-networks were notified of different or unresolved vulnerabilities than the ones in notifications to publishers, guaranteeing confidence in the results we observe. 
We believe these approaches effectively addressed the challenge of interdependent multi-stakeholder notification research.

\para{Source of activist notifications.}
We sent our activist notifications in collaboration with the CheckMyAds Institute, a well-known organization in the ad-tech community. Due to operational constraints, we could only use their branding on our emails, reports, and the website. We could not use their email servers to send notifications due to risks to their server reputation and potential interference with their operations.
In theory, this could have influenced how recipients perceived the email. To minimize this impact, we clearly stated CheckMyAds’ involvement in the opening sentence of the notification and used their branding prominently in the headers and footers of our emails, reports, and website. On a positive side, results in Section~\ref{sec:engagement:nature} suggests email branding to play a positive role in imposing sender branding. Some publishers and advertisers acknowledged being familiar with CMA's work, showing that email branding resulted in perception differences between activist and academic notifications---something that has not been studied before.

\para{Other limitations.}
The selection of control groups based on propensity matching does not account for factors like differences in technological capabilities, resources, or internal processes between treatment and control groups. We acknowledge this to be a limitation, however, we are limited by what we can observe as outsiders to the ecosystem and cannot account for these differences. Our approach  cannot identify dark pools stemming from obfuscated or encrypted advertising ID(s): also highlighted as a limitation of the approach that we follow~\cite{vekaria2023inventory}.

\subsection{Concluding remarks} \label{sec:discussion:implications}

\para{Alternative ordering of notifications.}
Beyond ordering explanation provided in \Cref{sec:methodology:notifications}, alternative ordering notifying the most influential entity before the final round (e.g., advertiser$\rightarrow$ad-network$\rightarrow$publisher, most-to-least influential) might resolve most dark pools in the first two rounds due to advertiser influence or pressure, but would make it statistically challenging to assess publisher impact later. 
The chosen ordering minimizes bias and isolates each notified entity’s impact on remediation. 
However, future work can delve deeper into exploring alternative ordering that mitigates potential risks by either relying on historical or archival analyses of webpages or leveraging previously observed incidents to act as a different kind of ``control'' in other multi-stakeholder settings.

\para{Multi vs. single stakeholder notifications.}
Notifying just the highest remediating stakeholder (e.g., ad-networks in our research) via single-stakeholder notifications would have still yielded sub-optimal results. 
First, it is often not known apriori that which stakeholder would be the highest remediating in different multi-stakeholder ecosystems. 
Second, in ecosystems with interdependencies between stakeholders where incentives are mis-aligned, single-stakeholder notifications may be effective for remediation, however, influence from other stakeholders can lead to \textit{more effective} remediation. 
Hence, notifying even just the most influential stakeholder would have proven to be sub-optimal as the \textit{most influential} stakeholder (i.e., advertisers in this case) might not be the \textit{most capable} stakeholder to resolve the vulnerability (i.e., ad-networks).

\para{Implications for online advertising.}
Our research illuminates how the online advertising supply chain functions and how different stakeholders interact. 
Our qualitative analysis shows that notifications can raise awareness about vulnerabilities in the supply chain, leading to increased transparency, trust, and the termination of relationships with problematic actors.
While notifications can effectively address vulnerabilities like dark pools in the online advertising ecosystem, there are important nuances. The motivations and capabilities of notification recipients play a crucial role. 
We found that publishers have the lowest dark pool remedial rates among stakeholders. In contrast, ad-networks, with their capability and resources, and advertisers, with their monetary incentives, are significantly more effective at resolving dark pools.
Additionally, we found that the source of the notification, whether from activists or academics, does not affect dark pool resolution, except for publishers.
Future work can investigate other notification channels (e.g., public naming and shaming \cite{tingley2022effects}) and external stakeholders (e.g., direct consumer notifications \cite{ahmad2024companies}).

\para{Operationalizing our notification pipeline to combat industry-scale ad inventory fraud}
Our findings provide a clear operational strategy for industry-wide, real-time notification feasibility. 
IAB Tech Lab---that introduced ad-tech supply chain transparency standards---already crawls \texttt{ads.txt} and \texttt{sellers.json} files on a daily-basis for automated validation checks via their Transparency Center~\cite{iab_transparency_center}.
They are in the best position to directly integrate our vulnerability detection logic based on their crawled files and by incorporating a dynamic analysis component to automatically flag and notify entities whose inventory relationships appear invalid or indicative of dark pooling. 
As IAB already maintains industry-wide mapping of inter-organizational relationships across both major and long-tail ad-tech participants, it can fully automate the validation and remediation loop. 
Furthermore, since the number of ad-networks---found to be the most influential stakeholders at mitigating dark pooling---is orders of magnitude smaller than the millions of publishers and brands, the IAB can easily maintain accurate contact channels such as verified email addresses for these entities, enabling targeted, high-impact notifications at scale. 
In combination, these capabilities and our research enables IAB to deploy a continuous, industry-level fraud-prevention pipeline that is more comprehensive and effective than any isolated remediation effort.

\para{Implications for other supply chains.}
Our results highlight the need to consider the complexities of multiple stakeholders in other ecosystems, such as the software supply chain, which includes operators, administrators, vendors, and developers. 
This need is further highlighted by Zimmermann et al. \cite{Zimmermann2019} who showed that vulnerabilities in widely-used open-source libraries impacted numerous software products. Unfortunately, the current state-of-the-art in security and privacy vulnerability notification only considers singlular stakeholders.
Similarly, in fintech, Abdou et al. \cite{Abdou2020} showed that inadequate API security could expose critical consumer financial data, impacting multiple stakeholders including consumers, banks, developers, and regulators in open banking ecosystem. 
Our work provides a model for testing the effectiveness of vulnerability notifications in these multi-stakeholder ecosystems.
When considering multi-stakeholder notifications in other ecosystems, we recommend identifying the most influential and the most capable stakeholders, along with the incentive flows and interdependencies within the ecosystem. 
Our work guides recommendation to (1) notify the most capable stakeholder prior to most influential one, and (2) the most interdependent stakeholder before the less interconnected ones. 
This is in order to create a build-up of pressure on the entities notified in latter stages.

\para{Multi-stakeholder notification research warrants more attention.}
Research on multi-stakeholder supply chains is complicated due to stakeholder inter-dependencies, existence and fixing capabilities of vulnerabilities across stakeholders and with mixed incentives. 
These complexities raise questions about which stakeholder to notify and how to notify each one. 
Our study, using the online advertising supply chain as a case-study, shows that not all stakeholders are equally capable or willing to resolve vulnerabilities. Furthermore, changing notification sources only, significantly influences some stakeholders.
This research is the first to tackle this scenario and highlights the need and aids examination of other fragmented multi-stakeholder ecosystems.

\section{Ethical Considerations}
\label{sec:discussion:ethics}

\para{IRB review.} Our institutional review board (IRB) reviewed this notification study and deemed it as not involving human subjects. Despite this, we designed and conducted our measurements and notifications following the principle of beneficence outlined in the Menlo Report \cite{MenloReport2012} and Belmont Report \cite{BelmontReport1979}. We aimed to maximize benefits and minimize potential harms.

\para{Infrastructure costs and risks.} 
We measured ad-tech supply vulnerabilities through crawls. To avoid stressing web servers, we did not conduct dynamic crawls of problematic publishers concurrently. We spaced out our additional crawls for \ads and \sellers files between 12 days to one month apart. Our crawlers did not follow the robots.txt directives on problematic publishers, but our methodology aligns with ethical and legal considerations for such audits \cite{sandvig2014auditing}. We sent notifications from a dedicated email server, set up with our university’s IT staff, to protect university mail server's reputation. 
We did not send repeated probes or marketing-style messages. The content was minimal---not the actual notification content, but just one sentence---asking to schedule a zoom call and we received no positive responses, ensuring negligible inbox burden on the recipients.

\para{Participation and Privacy risks.} 
We neither collect or record any personal information in this research nor include any organization names in the paper. Although we had phone calls and virtual meetings with several organizations to help them remediate vulnerabilities, we did not record or share any conversations or emails with anyone other than the authors. Moreover, we respected opt-out decisions of participants and deleted all their data. Using a codebook to code responses helped us prevent any deanonymization of the source of the response. We are committed to not sharing the actual mapping of the ad-tech entity name and their responses. We described our study to be part of a research study, disclosed our affiliations, and clearly outlined potential benefits and risks from the reported vulnerabilities in our notifications as well as on the project website shared with the recipients.

\para{Advertising costs.}
To understand which brands advertise on problematic publishers and which ad-networks are responsible, we clicked on the ads shown during page loads. 
The costs associated with our 28,376 ad clicks across 12 crawls of problematic publishers represented by label C and C’ in Figure~\ref{fig:timeline} of the paper were relatively minor, as per-ad CPMs are low, particularly for browsers without profiles \cite{olejnik2013selling, cook2020inferring}. 
These costs are justifiable given the benefits of understanding and remedying supply-chain vulnerabilities in the ad-tech ecosystem.
Quantitatively, 28,376/12 = 2365 ad clicks were performed per crawl distributed across 684 problematic publishers --- 2365/684 $\sim$3.5 ad clicks per publisher on an average. The average CPC costs for the open programmatic ad-tech ecosystem is \$0.63 for 2024~\cite{google_cpc_rates_2024} (as estimated for the dominant player – Google). So, the maximum revenue benefit our study could have caused using the average case is 3.5 * 0.63 = \$2.2 per problematic publisher.

\para{Privacy issues in online advertising.} 
Online advertising supports many web services. While many argue that the ad-tech ecosystem engages in questionable privacy practices (and we agree), we believe that approaches for improving its safety/security must be explored as it benefits users and stakeholders alike.

\para{Delayed notification to Control entities} 
We adopted delayed notification for control entities by notifying them at the end of the study only because it was the minimal-risk design that allowed sound attribution of the remediation effects. 
No alternative could have produced valid results without substantially increasing methodological (\Cref{sec:methodology:notifications}) and ethical risks. 
Notifying controls earlier---either at the time vulnerabilities were first computed or immediately after each round---would have collapsed the distinction between treatment and control.
This is because notifying control entities could have resulted in them reaching out to other ad-tech entities for resolution, making it difficult to understand if the remediation observed in the current as well as the following round was due to treatment notification or due to potential poisoning of some highly interconnected control entities (via business relationships) with treatment entities.
At the same time, the actual harms from this delay were objectively minimal. 
For publishers and ad-networks, dark pooling creates no monetary loss, and reputational risk is inherently constrained because adversaries disperse misuse across many IDs to avoid detection; for advertisers---the highest-risk group---we explicitly placed them in the shortest delay window (i.e. the last notification round), thereby minimizing exposure. 
From an external point-of-view, quantifying precise harm is infeasible without acces to the internal logs. 
However, all available evidence---including the very small number of vulnerable IDs per entity and the absence of any negative feedback from notified controls---indicate that any temporary risk was limited. 
Critically, our design adheres to Belmont~\cite{BelmontReport1979} and Menlo~\cite{MenloReport2012} principles: (1) it maximizes beneficence by enabling the first rigorous, multi-stakeholder evaluation of vulnerability notification efficacy in order to build automated remediation pipelines that will protect orders of magnitude more entities. (2) it distributes burdens fairly across different stakeholder types. 
Delayed controls are standard in randomized trials across disciplines---often with far higher stakes---and are accepted because they are the only way to produce generalizable knowledge. 
Consistent with those norms, we additionally mitigated residual risks by providing control entities with detailed guidance, responsive support, and post-study assistance via multiple email correspondance and/or virtual calls to ensure that they can fix the issues or clarify doubts with us throughout the process.
Thus, delayed notification was the ethically preferable and methodologically necessary choice, posing only minor, temporary risks while enabling long-term, large-scale harm reduction that no other study design could have achieved.


\ifCLASSOPTIONcompsoc
  \section*{Acknowledgments}
\else
  \section*{Acknowledgment}
\fi
\ifanonymous
This work was supported in part by the National Science Foundation
under award numbers 2138139, 2103439, and 2338377.
\else
\input{acknowledgment}
\fi

\bibliographystyle{IEEEtran}
\bibliography{mybib}

\appendices

\newpage
\section{Data Availability}
We will open source our dark pooling vulnerability detection and notification pipeline upon acceptance of this work. We will also include our notification template, vulnerability remediation guidance shared with the vulnerable entities and the website we used for the notified entities to access this information. We hope this will foster future notification studies as well industry initiatives such as IAB to incorporate such notification pipelines at larger scale.

\section{Example: ads.txt, sellers.json, Dark Pooling}
\begin{figure}[h]
    \centering
    \includegraphics[width=1\linewidth]{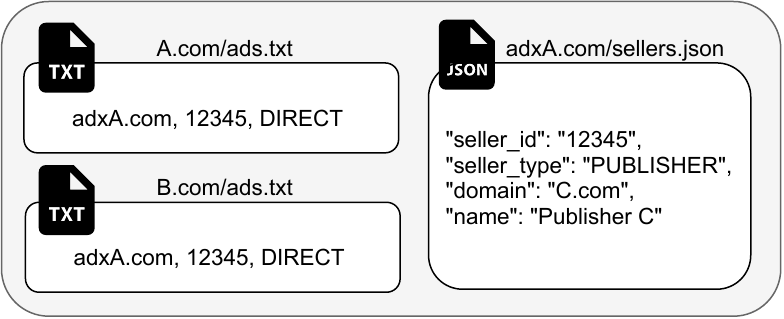}
    \caption{An example depicting the structure of entries in hypothetical \texttt{ads.txt} and \texttt{seller.json} files. In this example, publishers A, B, and C are considered unrelated (i.e., are not organizational related) and represents a scenario of dark pooling. The sellers.json of adxA.com shows that the seller account 12345 is owned by PublisherC. However, ads.txt of A.com and B.com lists seller account 12345 on adxA.com as their direct seller. This suggests dark pooling where A and B are pooling their ad inventories with C on adxA.com.}
    \label{fig:example-adstxt-sellersjson-darkpooling}
\end{figure}

Figure~\ref{fig:example-adstxt-sellersjson-darkpooling} represents structure of sample entries in \texttt{ads.txt} and \texttt{sellers.json} files. It also illustrates a typical dark pooling scenario.

\section{Categorization of Notification Responses}
Table~\ref{tab:thematic-responses} shows categorization of notification responses by sender reputation under different themes and thematic categories.

\begin{table*}[!t]
\vspace{-14cm}
\centering
\caption{Categorization of notification responses by sender reputation under different themes and thematic categories for victim publishers, ad-networks, and advertisers.} 
\begin{tabular}{@{}clcccccc@{}}
\toprule
\multirow{2}{*}{Theme} &
  \multicolumn{1}{c}{\multirow{2}{*}{Thematic Catgeories}} &
  \multicolumn{2}{c}{Victim Publishers} &
  \multicolumn{2}{c}{Ad Networks} &
  \multicolumn{2}{c}{Advertisers} \\ \cmidrule(l){3-8} 
                                               & \multicolumn{1}{c}{}                   & Activist & Academic & Activist & Academic & Activist & Academic \\ \midrule
\multirow{6}{*}{Approaches to Resolution}      & Actionable details requested           & 9        & 5        & 6        & 7        & 0        & 1        \\
                                               & Forwarded to the concerned team        & 7        & 10       & 0        & 0        & 5        & 0        \\
                                               & Collaborative resolution performed     & 3        & 2        & 4        & 4        & 0        & 1        \\
                                               & Fixed the reported issues              & 7        & 2        & 1        & 0        & 1        & 0        \\
                                               & Contacted responsible entities for fix & 3        & 3        & 0        & 2        & 0        & 1        \\
                                               & Justifying the reported issue          & 1        & 0        & 2        & 0        & 0        & 0        \\ \midrule
\multirow{6}{*}{Attitude towards Notification} & Acknowledged the notification          & 16       & 13       & 5        & 8        & 3        & 2        \\
                                               & Motivated to fix the issue             & 9        & 5        & 2        & 3        & 0        & 2        \\
                                               & Concerned about the notification       & 4        & 3        & 2        & 1        & 2        & 3        \\
                                               & Research initiative appreciated        & 4        & 1        & 3        & 1        & 0        & 0        \\
                                               & Awareness about the vulnerability      & 3        & 0        & 0        & 1        & 0        & 0        \\
                                               & Notification unhelpful or uninterested & 1        & 2        & 0        & 1        & 0        & 0        \\ \midrule
\multirow{4}{*}{Challenges to Resolution} &
  Doubting correctness of the report &
  5 &
  2 &
  4 &
  2 &
  0 &
  0 \\
                                               & Misinterpreted as phishing             & 2        & 1        & 1        & 3        & 1        & 1        \\
                                               & Limited resources to fully resolve     & 1        & 2        & 0        & 1        & 0        & 0        \\
                                               & Complete resolution beyond control     & 0        & 1        & 1        & 1        & 0        & 0        \\ \bottomrule
\end{tabular}
\label{tab:thematic-responses}
\end{table*}

\newpage
\section{Propensity-matched DiD Algorithm}
\label{appendix:PMDiD}

\noindent Algorithm 1 describes the pseudocode for performing propensity-matched difference-in-differences (PM-DiD).

\noindent\textbf{\rule{\linewidth}{1mm}} 
\begin{center}
\noindent \textbf{Algorithm 1:} Propensity Matched Difference-in-Differences
\end{center}
\noindent\textbf{\rule{\linewidth}{1mm}} 

\begin{algorithmic}[1]
\State \textbf{Input:} 
\State \hspace{4mm} Treatment group $\mathcal{T}$ (i.e., $\mathcal{T}_1$, $\mathcal{T}_2$ or $\mathcal{T}_1 \cup \mathcal{T}_2$)
\State \hspace{4mm} Control group $\mathcal{C}$
\State \hspace{4mm} Measure $M$
\State \textbf{Output:} 
\State \hspace{4mm} Effect size $\Delta_{(t,c)}$ 
\State \hspace{4mm} Statistical measures $N_{\text{rem.}}, \mu_{\text{ov.}}, \mu_{\text{rem.}}$
\vspace{1mm}
\Statex \textbf{Rounds and Measures:}
\State \hspace{4mm} Round 1: $M \gets probdomains_{vp}$
\State \hspace{4mm} Round 2: $M \gets pools_{ad}$ and $partnerpools_{ad}$
\State \hspace{4mm} Round 3: $M \gets pools_{ad}$
\Statex \textcolor{blue}{\rule{\linewidth}{0.5mm}}
\Statex \textbf{Step 1: Propensity Matching for each round}
\ForAll{$t \in \mathcal{T}$}
    \State Compute measure: $M^{t}_{\text{before}}$
    \State Find nearest neighbor $c \in \mathcal{C}$ such that:\\ 
    \hspace{8mm} $\min(|M^{t}_{\text{before}} - M^{c}_{\text{before}}|)$
    \State Add matched pair $(t, c)$ to matched pairs $\mathcal{P}$.
\EndFor
\vspace{1mm}
\Statex \textcolor{blue}{\rule{\linewidth}{0.5mm}}
\Statex \textbf{Step 2: Measure evaluation before and after intervention}
\ForAll{$(t, c)\ pairs \in \mathcal{P}$}
    \State Record pre-intervention measure values: $M^{t}_{\text{pre}}$, $M^{c}_{\text{pre}}$
    \State Record post-intervention measure values: $M^{t}_{\text{post}}$, $M^{c}_{\text{post}}$
    \State Compute effect size per pair:
    \Statex \hspace{5mm} $\Delta_{(t,c)} = (M^{t}_{\text{post}} - M^{c}_{\text{post}}) - (M^{t}_{\text{pre}} - M^{c}_{\text{pre}})$
\EndFor
\vspace{1mm}
\Statex \textcolor{blue}{\rule{\linewidth}{0.5mm}}
\Statex \textbf{Step 3: Statistical Metrics}
\State Initialize counters: $N_{\text{rem.}} \gets 0$, $\mu_{\text{ov.}} \gets 0$, $\mu_{\text{rem.}} \gets 0$
\ForAll{$(t, c)$ pairs}
    \If{$\Delta_{(t,c)} < 0$}
        \State Increment $N_{\text{rem.}} \gets N_{\text{rem.}} + 1$
        \State Update $\mu_{\text{rem.}} \gets \mu_{\text{rem.}} + \Delta_{(t,c)}$
    \EndIf
    \State Update $\mu_{\text{ov.}} \gets \mu_{\text{ov.}} + \Delta_{(t,c)}$
\EndFor
\State Compute final metrics:
\Statex \hspace{5mm} $\mu_{\text{ov.}} = \frac{\mu_{\text{ov.}}}{|\mathcal{T}|}$
\Statex \hspace{5mm} $\mu_{\text{rem.}} = \frac{\mu_{\text{rem.}}}{N_{\text{rem.}}}$
\vspace{1mm}
\Statex \textcolor{blue}{\rule{\linewidth}{0.5mm}}
\State \textbf{Return:} $\Delta_{(t,c)}, N_{\text{rem.}}, \mu_{\text{ov.}}, \mu_{\text{rem.}}$
\end{algorithmic}

\noindent\textbf{\rule{\linewidth}{1mm}} 

\end{document}